\documentclass[11pt,twoside]{article}
\usepackage{asp2010}

\resetcounters

\begin{document}

\title{Mid-Infrared Variation in Young Stars}
\author{L.\ M.\ Rebull$^1$} 
\affil{$^1$Spitzer Science Center, MS 220-6, 1200 E. California Blvd.,
Pasadena, CA, 91125, USA}

\begin{abstract}
Since 2003, the Spitzer Space Telescope has provided groundbreaking
views of Galactic star formation in bands from 3.6 past 24 microns. 
During the cryogenic mission (the first 5.5 years), variability of
young stars at these bands was noted, although typically with just a
few epochs of observation. The cryogen ran out in 2009, and we are now
in the warm mission era where the shortest two bands (3.6 and 4.5
microns) continue to function essentially as before.  The phenomenal
sensitivity and stability of Spitzer at these bands has enabled
several dedicated monitoring programs studying the variability of
young stars at timescales from minutes to years.  The largest of these
programs is YSOVAR (Stauffer et al.), but there are several smaller
programs as well.  With at least as many as 2200 young star light
curves likely to come out of this, these programs as a whole enable
more detailed study of the young star-disk interaction in the infrared
for a wider range of ages and masses than has ever been accomplished
before. Early results suggest a wide variety of sources of
variability, including dust clouds in the disk, disk warps, star
spots, and accretion.  This contribution will review some of the most
recent results from these programs. 
\end{abstract}

\section{Overview of young stars}

The general outline of the formation of low-mass stars has been widely
accepted for at least 20 years (see, e.g., Bertout 1989). An initial
molecular cloud collapses onto itself, forming an envelope and then a
disk around a central mass; jets help regulate angular momentum in the
early phases and perhaps the interaction of the magnetic field with
the circumstellar disk regulates the angular momentum in later stages
(e.g., K\"{o}nigl 1991, Shu {\it et al.}~2000). 

\begin{figure}[!ht]
\plotone{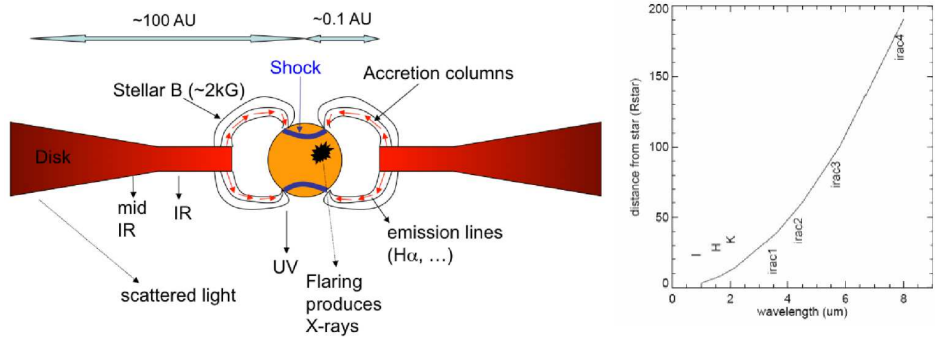}
\caption{LEFT: Anatomy of a young star, after Hartmann (1998). Not to scale!
Note that the mid-infrared emission comes from relatively far out in
the disk. RIGHT: Simplified version of relationship between peak emission
wavelength and distance from the protostar.  For a typical low-mass
protostar, the near-infrared bands $I$, $H$, and $K$ sample within
$\sim$10$R_*$; the four IRAC bands at 3.6, 4.5, 5.8, and 8 $\mu$m are
sensitive to disk properties at $\sim$30--$\sim$200$R_*$.}
\label{fig:anatomy}
\end{figure}

Figure~\ref{fig:anatomy} shows the basic ``anatomy'' of a young
stellar object (YSO), at ages of $\sim$1-5 Myr, when there is still a
substantial circumstellar disk but no envelope or jets. In this
Figure, the circumstellar disk is flared at the outer edges, and the
inner edge is truncated by the protostellar magnetic field. The
completely convective young star is rotating quickly, and as such has
a strong magnetic field. Accreting matter follows the field lines, and
crashes onto the protostar near the magnetic poles.  The active young
star produces flares in X-rays, ultraviolet from the accretion shocks,
emission lines from the accretion columns, and infrared from the disk
itself. Near-infrared (NIR) emission originates closer into the
central object than mid-infrared (MIR).  Note that even in this simple
picture, very few of these properties are likely to be constant even
over relatively short time intervals; rotation, accretion, flares, and
even inhomogeneities forming and dispersing in the disk are all highly
dynamic processes. 

Figure~\ref{fig:anatomy} also shows (on the right) a schematic, simple
approximation for the relationship between peak emission from the disk
and distance from the central protostar.  Some protostars have disk
emission starting at wavelengths as short as the NIR $JHK$, 1-2
$\mu$m; these disks likely are quite close in to the central object,
on the order of $\sim$20$R_*$. However, in the MIR (such as the
Spitzer Space Telescope bands at 3.6-8 $\mu$m), we sample disk
properties much further out, from $\sim$30$R_*$ to $\sim$200$R_*$.  In
reality, this is a vast simplification, and heated inner disk walls
and/or rims, system inclination, disk-photosphere contrast, and many
other properties in addition to the temperature of the central object
affect what location in the disk a given wavelength samples.  In the
relatively extreme case of HH 30, a nearly edge-on disked young star
studied with the Hubble Space Telescope in the optical and NIR,
indications can be seen of a light beam (or shadow?) from the central
source sweeping across the flared disk with a period of $\sim$7.5d 
(Duran-Rojas {\it et al.}~2009, Watson \& Stapelfeldt 2007). Reality
is complicated.

In the rest of the contribution, I will attempt to address whether
young stars really do vary in the MIR, and if so, on what timescale. 
An important next step in understanding any variability is determining
whether the source of any MIR variation is really at $\sim$10s of
$R_*$ or at some other significantly different distance.

\section{MIR YSO variability in the pre-Spitzer era}

Prior to the advent of the Spitzer Space Telescope (Werner {\em et
al.}~2004), there were occasional published references to variability
in YSOs in mid-infrared wavelengths, such as the following.

Prusti \& Mitskevich (1994) originally set about looking for
variations in all the repeated observations of Herbig AeBe (HAeBe)
stars found at 12 and 25 $\mu$m in the Infrared Astronomy Satellite
(IRAS) data taken in 1983.  However, they found that source confusion
was prohibitive, and focused their study on two HAeBe stars, AB Aur
and WW Vul.  They found significant variations on timescales ($t$) of
months. They suggested that cometary clumps or a clumpy wind were
plausible explanations for the variations observed.

Liu {\em et al.}~(1996) reported that they found MIR variations in
their ground-based data, ranging in amplitude from 30-300\%, on
timescales of days to years. They pointed out that the MIR variations
most likely do not have the same origin as the optical/NIR variability.
They suggest that since most of the MIR is from
disk, then the cause of the variability must be there too.  In order
to achieve the variations that they observed, they postulate that the
mass accretion rate ($\dot{M}$) varies by an order of magnitude. 
Small-amplitude changes in the MIR could be due to reprocessed
accretion luminosity, whereas larger changes could be due to disk
accretion rate, a disk instability, or outflow activity.

Abraham {\em et al.}~(2004) took on the relatively difficult task of
comparing Infrared Space Observatory (ISO) data taken in 1995-98 to
MIR data taken at other bands with other facilities/instruments (such
as MSX) at other epochs. They studied 7 FU Ori objects, and found weak
MIR variability on timescales of years (over 1983--2001).

The next year, Barsony {\em et al.}~(2005) reported on ground-based
observations in the MIR of embedded objects in the $\rho$ Ophiuchi
cloud core.  By comparison to ISO data, they found significant
variability in 18 out of 85 objects detected, on timescales of
years.   They found such variability in all spectral energy
distribution (SED) classes with optically thick disks, and suggest
that this might be due to time-variable accretion.

Later that year, in a large paper covering the MIR properties of the
Orion Nebula, Robberto {\em et al.}~(2005) reported on MIR variability
in Orion, almost as an afterthought. They found variations up to
$\sim$1 mag, on timescales of $\sim$2 years.  They invoke changes in
$\dot{M}$, activity in the circumstellar disk, or changes in the
foreground $A_v$ to explain the variations they see.

Finally, Juhasz {\em et al.}~(2007) report on the ISO variability of 
SV Cep.  SV Cep is a UX Orionis-type variable, the generic properties
of which include intermediate-mass YSOs with short ($t\sim$days-weeks)
eclipse-like events in the optical. These could be edge-on
self-shadowed disks, for example.  This study is the only one (at
least, the only one of which we have knowledge) reporting on a
monitoring campaign conducted with ISO itself (as opposed to
comparison of ISO data to data taken with other
instruments/facilities).  They obtained contemporaneous optical
monitoring data over ISO's lifetime (1995--1998) to aid in the
interpretation of the MIR light curves.  They found significant MIR
variability on $t\sim$25 months; the MIR variations were
anti-correlated with the optical variations but the far-IR variations
were correlated with optical.  They suggest a self-shadowed disk with
a puffed-up inner rim, but find that this model does not do well at
reproducing the MIR variations; again, $\dot{M}$ variations are
invoked to explain the MIR variations.

\section{Results in the Spitzer era}

\subsection{Introduction to Spitzer}

The Spitzer Space Telescope (Werner {\it et al.}~2004) is an 85 cm,
f/12 telescope.  Before the on-board cryogen was exhausted, it
operated at $<\sim$5.5 K, and was background-limited at 3-180 $\mu$m.
It has two science cameras (Infrared Array Camera -- IRAC -- Fazio
{\it et al.}~2004 and the Multiband Imaging Photometer for Spitzer --
MIPS -- Rieke {\it et al.}~2004), plus a low/moderate resolution
spectrograph (Infrared Spectrograph -- IRS -- Houck {\it et
al.}~2004).  Launched August 2003 into an Earth-trailing orbit, it was
10-1000 times more sensitive than the 1983 IRAS mission. 

The cryogen ran out in May 2009, and the telescope passively remains
at $\sim$30 K.  At this temperature, the IRAC 3.6 and 4.5 $\mu$m
channels still operate essentially as they did before cryogen
exhaustion, which is still 120-1000 times faster than VLT or Keck. 
This portion of the mission is ``Spitzer-Warm'', and NASA has
committed to fund $\sim$3 years  of warm operations.  As part of the
Warm Mission, large ($>$500 hours), coherent observing programs were
solicited, called ``Exploration Science'' programs.

The cryogenic Spitzer legacy for star formation research is
substantial.  There are multi-band maps of $\sim$300 square degrees of
the Galactic plane, with $>$100 million sources. There are maps of
$\sim$70 square degrees in nearby ($d<$500 pc)  star-forming regions,
with $\sim$8 million total sources in Taurus, Ophiuchus, Perseus,
Chamaeleon, Serpens, Auriga, Cepheus, Lupus, Orion clouds, etc. 
Conservatively, we estimate that there are $\sim$20,000 YSOs in this
rich data set.  

Spitzer is a superb telescope for photometric monitoring because it is
stable (better than 1\%) and sensitive, wide-field (a single IRAC
field of view is 5$^{\prime}$ on a side), Earth-trailing (so no
orbital day/night aliasing), and it observes at bands sensitive to
both photospheres and dust.  In the Warm Mission era, we have the same
amount of observing time as in the cryogenic era, and ``just'' 2
channels.  There are several Exploration Science and smaller programs
exploring the time domain with Spitzer.

%

\subsection{Variability at Spitzer bands}

YSO variability at Spitzer bands is unambigously  apparent, and the
torrent of papers on the subject is still ramping up.  In the below, I
discuss the papers in the order in which they appeared in the
published literature.

The Legacy program ``Cores to Disks'' (c2d; Evans {\it et al.}~2003,
2009) took two epochs of observation (both IRAC and MIPS) separated by
several hours to allow for asteroid removal. Several different papers
(Alcala {\it et al.}~2008 and references therein) looked for variation
between these two epochs (on timescales of $\sim$3-6 hrs), and did not
find anything believable (within $\sim$25\%) at wavelengths 3.6-24
$\mu$m. 

Another Legacy program, ``Surveying the Agents of a Galaxy's
Evolution'' (SAGE; Meixner {\it et al.}~2006) studied the Large
Magellanic Cloud (LMC), again in two epochs (both IRAC and MIPS), but
this time separated by $\sim$3 months. Vijh {\it et al.}~(2009) report
on all of the variables found by comparing these two maps. They found
mostly asymptotic giant branch (AGB) stars, which they point out is
not  entirely unexpected.  However, we note here that optical
variability is one of the defining characteristics of YSOs, and AGB
stars are the most common ``contaminant'' in Spitzer selection of YSO
candidates;  having the right MIR colors plus MIR variability does not
ensure that a given object is necessarily a YSO. Vijh {\it et
al.}~(2009) find 29 massive (=HAeBe) YSO candidates out of nearly 2000
variables, which they interpret to mean that at least 3\% of all
massive YSOs are variable. They also report that the amplitude of
variability is often greatest at 24 $\mu$m, perhaps because most of
their YSO SEDs peak at 24 $\mu$m (or longer). 

The first high-cadence monitoring of young stars in IRAC bands was
conducted by Morales-Calder\'on {\it et al.}~(2009).  The stars in IC
1396A were monitored twice a day for 14 d, plus every $\sim$12 s for 7
hrs. More than half of the YSOs showed variations, from $\sim$0.05 to
$\sim$0.2 mag, on a wide variety of timescales, which enables the
first possible serious physical interpretations of the variations.
About 30\% of the YSOs had quasi-periodic variations, on timescales of
$\sim$5-12d periods, which they interpreted as  1 or 2 high-latitude
spots illuminating inner wall of the circumstellar disk, plus a large
inclination angle. Two objects have variations on timescales of
$\sim$hours, but no color in the variations, which is interpreted as
flares, and/or possibly $\dot{M}$ flickering. Other light curves are
more likely due to varying $\dot{M}$ or disk shadowing. About 20\% of
the IC 1396A YSOs vary on $t\sim$days, without color changes, which
could be due to $\dot{M}$ variations, and/or rapidly evolving spots. 
There are three objects that vary on timescales of days, with color
variations, which the authors interpreted as radial differential
heating of the inner disk, and possible inner disk obscurations. 
There were 46 variables not  identified as YSOs (e.g., without a
discernible IR excess); possibly they are YSOs or even AGBs, but more
data are needed to interpret these. Larger amplitude variables tend to
also be more embedded objects, but an order of magnitude change in
$\dot{M}$ is needed to match the light curves, so this is probably not
the dominant factor. A simple starspot is insufficient to explain the
variability, but a hotspot combined with disk inhomogeneities does
work.  Also in the data was a young $\delta$ Scuti star, with a 3.5 hr
periodicity on top of a $\sim$9 d period.

\begin{figure}[!ht]
\plotone{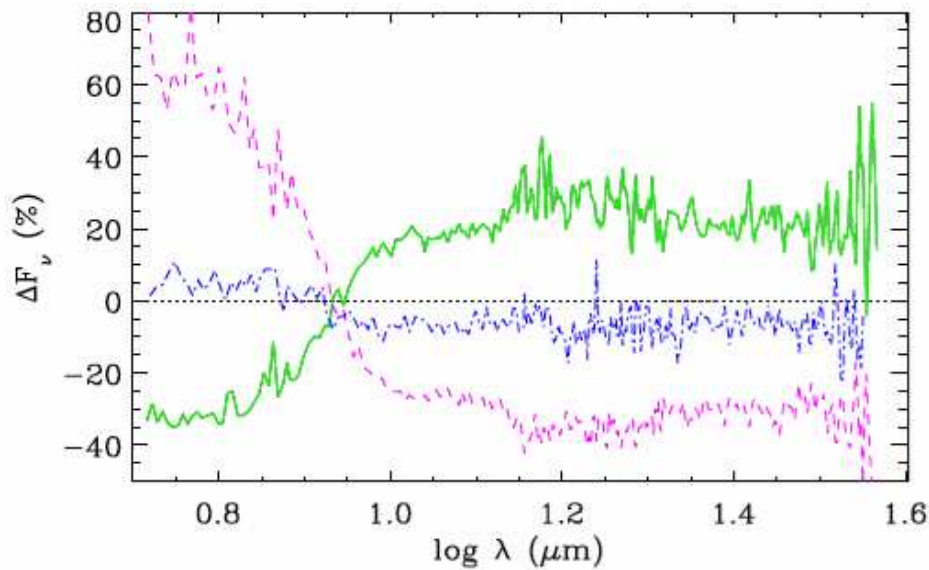}
\caption{Figure 1b from Muzerolle {\it et al.}~(2009).  Difference
spectra between the first and second epochs (solid green), second and
third epochs (dashed magenta), and third and fourth epochs (dash-dot
blue), as a percentage change in flux. First epoch was 2007 Oct 9,
second epoch was 2007 Oct 16, third epoch was 2008 Feb 24, and fourth
epoch was 2008 Mar 2. Variations ``pivot'' at $\sim$ 8.5 $\mu$m, and
are as large as 20-30\% in a week.}
\label{fig:james}
\end{figure}

Working in IC 348, Muzerolle {\it et al.}~(2009) report specifically
on the variations they observed in the T~Tauri star LRLL 31. This
object is identified specifically as a ``transition disk'', meaning
that it falls in a category of object thought to be in transition
between a primordial, thick disk and a disk actively forming planets
with gaps and structure in the disk created by protoplanets. The SED
for this object suggests a large inner hole or gap. Muzerolle {\it et
al.}~(2009) initially noticed variations in IRAC+MIPS (3.6-24 $\mu$m) 
observations taken over $t\sim$5 months. They used both IRS and MIPS
(5-40 $\mu$m) to further probe these variations on timescales of
$\sim$days to $\sim$months.  In the IRS spectra, reproduced in
Figure~\ref{fig:james}, they found that the variations pivot at a
point $\sim$8.5 $\mu$m, and they found variations of 20-30\% within a
week. They also found variations at 24 $\mu$m on $t\sim$1 day; recall
Figure~\ref{fig:anatomy} -- note that variations on those timescales
are certainly not very far away from the star, even at that
wavelength.  Muzerolle {\it et al.}~(2009) interpret these
observations as vertical variations of an optically thick annulus
located close to the star. Variations in $\dot{M}$ (up to a factor of
5) could be contributing here, or a companion causing gap, or even a
warp.

Giannini {\it et al.}~(2009) conducted observations of the Vela
Molecular Ridge (VMR-D), with just IRAC.  The two maps were taken
$\sim$6 months apart. They simply accept variability of YSOs at the
MIR bands as a defining characteristic of YSOs, as a statement of
fact, and do not attempt to further justify it.  This suggests a
change in culture in the community. Giannini {\it et al.}~(2009)
conclude that 19 (out of $\sim$200) are likely variable young stars.

Bary {\it et al.}~(2009) obtained IRS spectra of 11 actively accreting
T~Tauri stars in Taurus-Auriga; 2 of the 11 (DG Tau and XZ Tau) had
significant variation in the 10 $\mu$m silicate feature on timescales
pf $\sim$months to years (not weeks). They point out that this
timescale is consistent with the source of the variations being
motions of dust in the disk at $R<\sim$1 AU, and not with a clumpy
dust envelope. Disk shadowing could still be possible, especially at
the longer timescales. The possibility remains that there are binary
companions to these objects as well. They had difficulty in fitting
the line profile with existing models, suggesting that similar
problems encountered by other investigators fitting single-epoch
observations of other sources may ultimately be due to similar
time-dependencies in those other sources. In any case, vertical mixing
and disk winds are likely to be significant components of the source
of the variability.

\section{YSOVAR}

John Stauffer leads the Exploration Science program (from Spitzer's
Cycle 6) entitled, ``Young Stellar Object Variability: Mid Infrared
Clues to Accretion Disk Physics and Protostar Rotational Evolution,''
or YSOVAR.  We were allocated 550 hours to conduct the first sensitive
MIR (3.6 and 4.5 $\mu$m) time series photometric monitoring of several
star-forming regions on timescales of $\sim$hours to years.  Our
fields include $\sim$1 square degree of Orion (centered on the Orion
Nebula Cluster) plus smaller $\sim$25 square arcminute regions in 11
other well-known SFRs: AFGL 490, NGC 1333, Mon R2, NGC 2264, Serpens
Main, Serpens South, GGD 12-15, L1688, IC1396A, Ceph C, and IRAS
20050+2070.  Details of our fields, as well as a complete list of our
collaborators, can be found at our website:
http://ysovar.ipac.caltech.edu.

For our observations, we typically obtain $\sim$100 epochs/region
(sampled $\sim$twice/day for 40d, less frequently at longer
timescales). We started obtaining data in Sep.\ 2009 and will be
obtaining data through June 2011.  At the completion of our program,
there should be good light curves for at least $\sim$2200 YSOs! We are
also obtaining simultaneous (or nearly simultaneous) ground-based
monitoring at $I_c$, $J$, and $K_s$, which aid significantly in our
ability to interpret the light curves.  (NB: if anyone in the
community is interested in helping obtain such data, please contact us
at ysovar-at-ipac.caltech.edu.)

Note that we include under the YSOVAR umbrella some affiliated
programs such as J.\ Stauffer's Cycle 7 Orion follow-up on some of our
targets discussed below, P.\ Plavchan's Cycle 6 Rho Oph intensive
monitoring, K.\ Covey's Chandra/Spitzer Ceph C monitoring, and J.\
Forbrich's GGD 12-15 Chandra/Spitzer monitoring. As of this writing,
there are five clusters with at least some data: Orion, L1688, Ceph C,
IC 1396A, IRAS 20050.

\begin{figure}[!ht]
\plotone{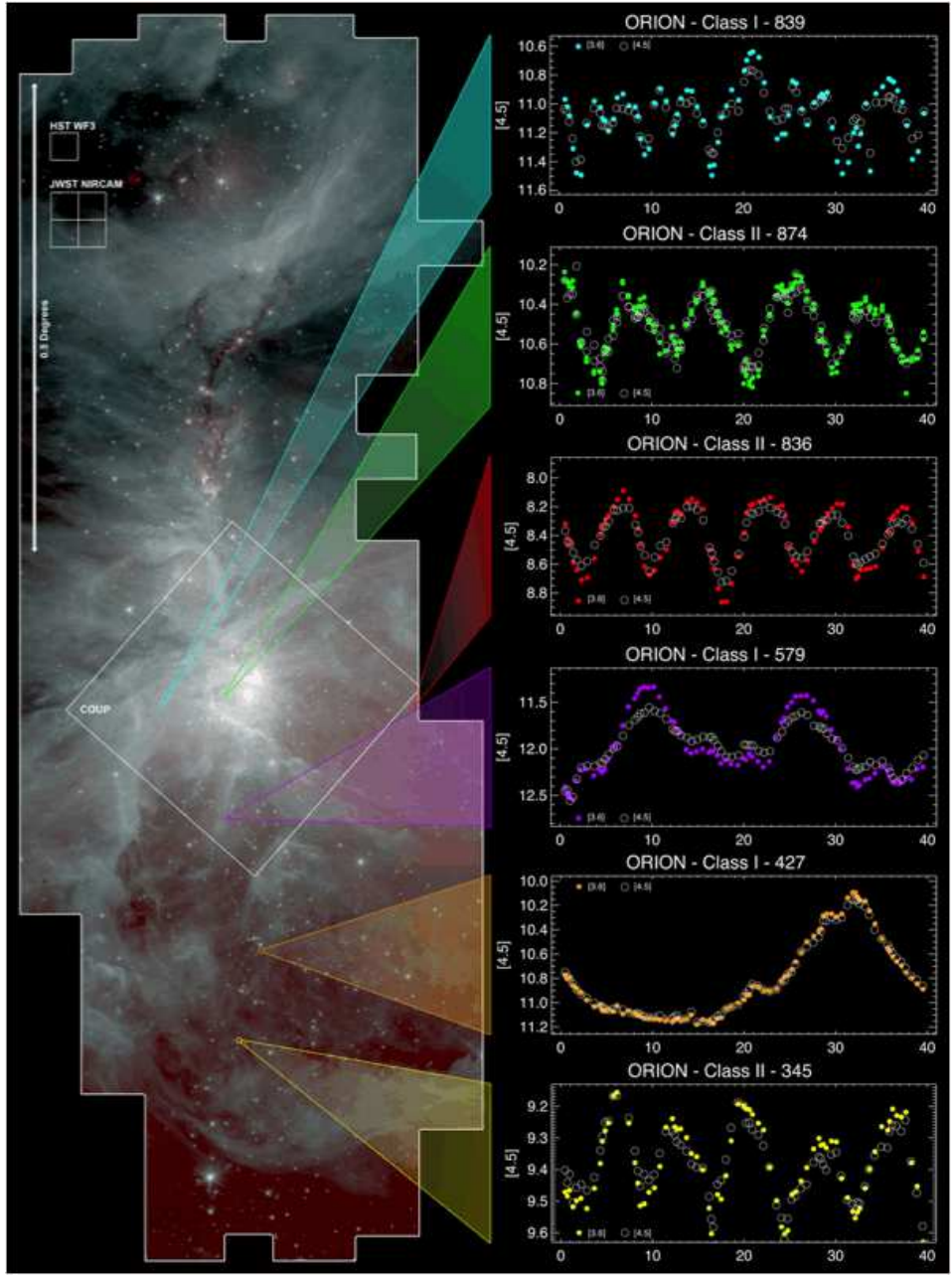}
\caption{Image depicting some of the variability found in YSOVAR
observations in Orion. The image on the left is the two-band (3.6 and
4.5 $\mu$m) Spitzer composite. Indications of the relative sizes of
0.5$^{\circ}$, and the Hubble WF3 and the JWST NIRCAM fields of view
are in the upper left.  The Chandra  COUP field is indicated, centered
on the Trapezium. For each of the light curves depicted, the solid
point is 3.6 $\mu$m and the hollow point is 4.5 $\mu$m. Note the
diversity of behavior exhibited by these variables.}
\label{fig:seniorrev}
\end{figure}

Morales-Calder\'on {\it et al.}~(2010, 2011; see also this volume)
report on the early results from the YSOVAR monitoring of Orion. We
find variability in $\sim$65\% of the objects with infrared excess
(Class I+II) and $\sim$30\% of the objects without infrared excess
(Class III). It should not be surprising that there is tremendously
diverse behavior exhibited by these variables.
Figure~\ref{fig:seniorrev} shows just a sample of some of the light
curves from some of the objects in Orion.  The shape of the light
curves likely have origins in slow changes in $\dot{M}$, changes in
the $\dot{M}$ geometry, flares, photospheric spots, disk warps, and
some causes yet to be identified! The contemporaneous optical and NIR
data sometimes have a similar shape and amplitude as the MIR light
curves, sometimes the NIR has a much larger amplitude, and sometimes
the NIR variations are much smaller or not variable at all. In some
cases, the NIR variations are phase-shifted with respect to the MIR.
(See Morales-Calder\'on {\it et al.}~2010 for example light curves and
more discussion.)

Because the emission in the MIR is likely coming from the disk
(thermal dust emission) as well as the photosphere, the variations we
see are likely due to variability in the disk as well as the
photosphere.  Thus, it is in general harder to derive a period for the
central YSO for our target objects than from light curves, say, in
$I_c$, where most of the emission comes from the photosphere (and
spots rotating into and out of view generate rotationally modulated
light curves).  For just 16\% of the variable objects
with infrared excess (Class I+IIs) can we derive a period, and most of
those are the ones with smaller excesses (90\% of those are Class IIs,
10\% are Class Is). For members without an IR excess, 40\% are
variables, and most of those are periodic. We can report $>$100 new
periods. Of the Orion members with period measurements in the
literature, we recover about 45\% of those.  There are
also 10 eclipsing binaries, 5 of which are new discoveries
(Morales-Calder\'on 2011). 

\begin{figure}[!ht]
\plotone{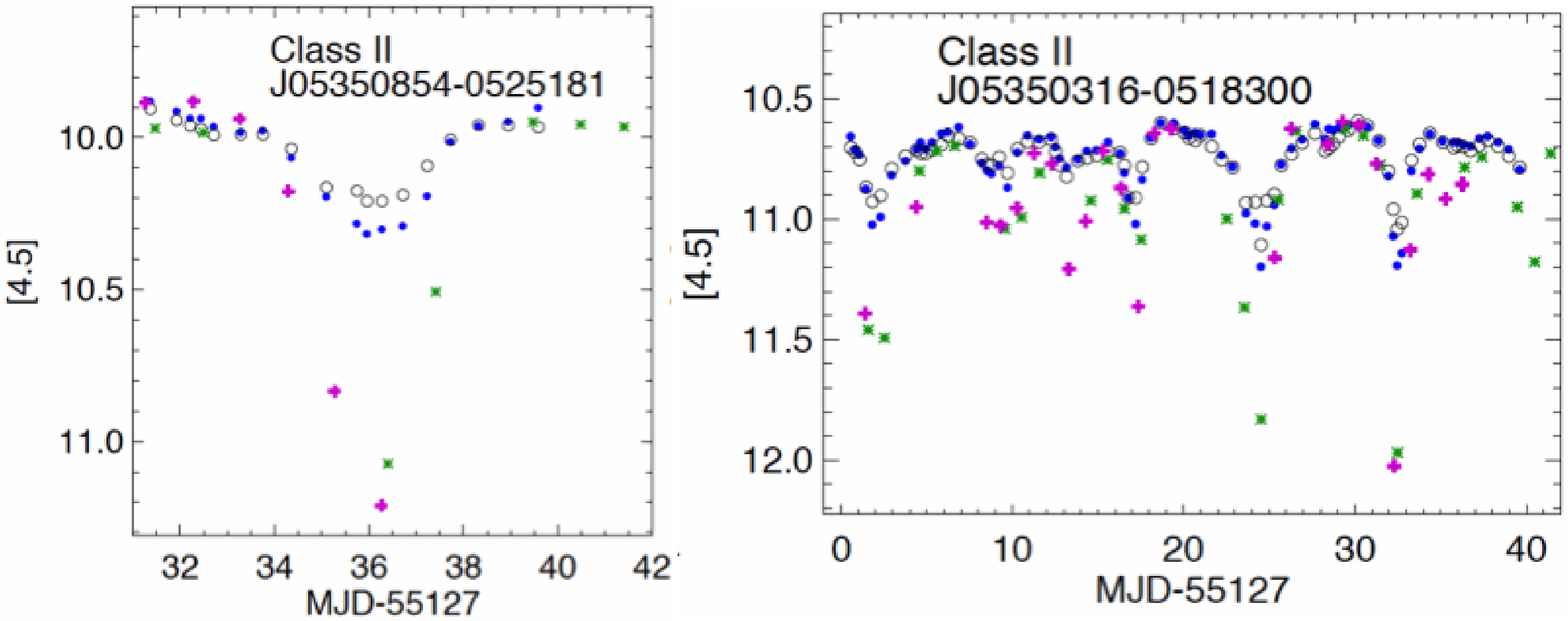}
\caption{Two examples of AA Tau-like (``dipper'') variability in our
Orion data, from Morales-Calder\'on {\it et al.}~(2010).  Solid blue
circles are 3.5 $\mu$m, open black circles are 4.5 $\mu$m, red or
green $*$ are $J$, and magenta $+$ are $I_c$. The light curves are
shifted in $y$-axis to align to the [4.5] ``continuum'' level.}
\label{fig:dipper}
\end{figure}

One significant class of variables that we have discovered have AA
Tau-like variations (see Bouvier {\it et al.}~2007 and references
therein for discussion of AA Tau). These ``dipper'' stars have  narrow
flux dips, on timescales of days, and typically more than one dip are
seen over our 40 d window; see Figure~\ref{fig:dipper}.  In order for
us to categorize a given object as a dipper, we require that the dip
is seen in more than one epoch unless there are corroborating data at
another band. Any optical or $J$ band corroborating data must have the
dips be  deeper by at least 50\%. The ``continuum'' of the light curve
must be flat enough that dip ``stands out.'' We find 38 Class I or II
objects ($\sim$3\%) in our set that are dippers, and we interpret this
variability  as structure in the disk, such as clouds or warps. 

Other upcoming results include the following. Plavchan {\it et
al.}~(priv.\ comm.) report that WL 4 is still eclipsing, 10 years
after the 2MASS calibration data (Plavchan {\it et al.}~2008) were
taken. This system is probably a quasi-stable disk eclipsing a binary
system like KH-15D.   Muzerolle, Flaherty {\it et al.}~(priv.\ comm.;
see also this volume) studied IC 348 and find IRAC variability (on
timescales of days to years) in 56\% of Class 0/I objects, 69\% of
Class II objects, and 58\% of the transition disks. Moreover, even at
24 $\mu$m, 60\% of Class 0/I, 40\% of Class II, and 40\% of transition
disks vary! They also find dips in the light curves like the YSOVAR
dippers.

The YSOVAR data set (as well as the associated programs) are certain
to yield interesting results in the coming years. For lack of space, I
have not addressed any possible monitoring results from Herschel or
WISE, much less any recent non-MIR monitoring of young stars, such as
CoRoT monitoring of NGC 2264 (see, e.g., Alencar {\it et al.}~2010 and
references therein for more information).

\section{Conclusions}

While 15 years ago, we were as a community uncertain as to whether
young stars vary in the mid-infrared, the literature suggested at
least small variations on timescales of months to years, likely due to
the disk. However, with the advent of the Spitzer Space Telescope and
its stable, sensitive, wide-field platform for monitoring young stars,
it has become unambiguous that yes, young stars vary in the
mid-infrared, and they vary on pretty much any timescale that one
cares to observe them (much as they do at many other bands).  While
definitive physical explanations for all of the tremendous diversity
of variability types is still elusive, strong candidates for some
types of variation are emerging. Some of the variability is clearly
due to photospheric spots, much is due to structure in the disk, some
is variation in mass accretion rate. Rotation, and the dynamic nature
of the young star-disk system, are both clearly important.  The
answers are still forthcoming!

\acknowledgements

I wish to acknowledge many helpful conversations with J.\ Stauffer,
M.\ Morales-Calder\'on, P.\ Plavchan, K.\ Covey, J.\ Carpenter, and
the rest of the YSOVAR team. I also wish to thank J.\ Muzerolle  for
pre-publication access to his results.  This work is based in part on
observations made with the Spitzer Space Telescope, which is operated
by the Jet Propulsion Laboratory, California Institute of Technology
under a contract with NASA. Support for this work was provided by NASA
through an award issued by JPL/Caltech.

\end{document}